# Comparing CSI and PCA in Amalgamation with JPEG for Spectral Image Compression

Muhammad SAFDAR,[1] Ming Ronnier LUO,[1,2] Xiaoyu LIU[1, 3]
[1] State Key Laboratory of Modern Optical Instrumentation, Zhejiang University, Hangzhou, China
[2] School of Design, University of Leeds, Leeds, UK
[3] College of Science, Harbin Engineering University, Harbin, China

**ABSTRACT**

Continuing our previous research on color image compression, we move towards spectral image compression. This enormous amount of data needs more space to store and more time to transmit. To manage this sheer amount of data, researchers have investigated different techniques so that image quality can be conserved and compressibility can be improved. The principle component analysis (PCA) can be employed to reduce the dimensions of spectral images to achieve high compressibility and performance. Due to processing complexity of PCA, a simple interpolation technique called cubic spline interpolation (CSI) was considered to reduce the dimensionality of spectral domain of spectral images. The CSI and PCA were employed one by one in the spectral domain and were amalgamated with the JPEG, which was employed in spatial domain. Three measures including compression rate (CR), processing time (Tp) and color difference CIEDE2000 were used for performance analysis. Test results showed that for a fixed value of compression rate, CSI based algorithm performed poor in terms of $\Delta E_{00}$, in comparison with PCA, but is still reliable because of small color difference. On the other hand it has lower complexity and is computationally much better as compared to PCA based algorithm, especially for spectral images with large size images.

## 1. INTRODUCTION

In last two decades, demand of high speed data transfer and high level of data compression have increased and it has been a hot topic of research in the area of signal processing and communications. As images are always a reasonable part of data that is being transmitted or stored, image compression techniques play an important role to compress such data. The visual information is compressed by human visual system and still it provides high quality of image and true color information. This information motivated researchers to compress visual data or images in a way that the compressed information should be reliable and color information should not change. Almost the same rate of compression was achieved in early 90's for the color images.

Concerning multi-spectral or hyper-spectral imaging which have a number of spectral bands, their accessibility is hindered by the communication bandwidth and image size. These limitations may be alleviated by efficient image compression. J. Delcourt et al., proposed an adaptive multi-resolution based multispectral imaging technique substituted with JPEG algorithm (Delcourt 2010). Markas and Reif proposed two coder schemes, one to remove spectral redundancy and another wavelet based coder to remove special redundancy, in the multispectral images (Markas 1993). N. Salamati et al., proposed a coder that puts a threshold on DCT components due to high correlation between visible color RGB and near-infrared components of four channel multispectral images. They also compared their results with JPEG and PCA based compression techniques and found almost the same compression rate (Salamati 2012). Canta and Poggi, proposed address predictive vector quantization based technique for multispectral images (Canta 1997). In this work both spatial and spectral dependencies were exploited. L. Chang used Eigen-based segmentation for multispectral image compression (Chang 2004). M. Cagnazzo, et al., proposed classified transform coding based spectral image compression technique in order to lower the computational complexity (Cagnazzo 2006). Q. Du and J. E. Fowler, amalgamated PCA with JPEG for hyper-spectral image compression (Du 2007).

The joint photographic experts group (JPEG) proposed an algorithm called JPEG which is the most commonly used image compression standard. By selecting a tradeoff between image quality and rate of compression, it can achieve a degree of compression that is desired depending on the application (Skodras 2001). The cubic spline interpolation (CSI) is a best polynomial based interpolation technique that is smooth on the edges due to its strict constraint of continuity at second derivative. It is a computationally simple and robust technique. It can be combined with JPEG for image compression.

In the current work, we continued our previous research on color image compression (Safdar 2014), we have incorporated cubic spline interpolation (CSI) into the JPEG algorithm to compress multispectral images. The CSI was used to compress spectral domain while JPEG was employed to reduce spatial dimensionality. The results of both methods were compared and analyzed on the bases of computational complexity and quality of the image reduced due to compression.

## 2. METHOD

The principle component analysis (PCA) has widely been used for spectral reduction and spectral de-correlation in multivariate data analysis (Du 2007). PCA practically provides optimal and excellent de-correlation in statistical sense. At the time of training of all PCs, PCA is commonly known as the KarhunenLo`eve transform (KLT), in which case a hyper-spectral image with N spectral bands produces an N×N unitary KLT transform matrix. Because of being data dependent matrix, it must be communicated to the decoder in any KLT-based compression system. Alternatively, corresponding to the P largest eigenvectors, by training in the KLT transform matrix PCA can effectuate dimensionality reduction. The data volume passed to the encoder then has P < N, rather than N, spectral components, and the resulting N×P PCA matrix is communicated to the decoder.

The JPEG 1992 standard was used in the current work. The PCA transform matrix and data mean vector were generated in MATLAB; the Kakadu encoder performs the spectral transform (implicitly reducing the spectral dimensionality if P<N) and then embeds the mean vector as well as the inverse transform matrix into the JPEG2000 bit-stream. The encoder automatically allocates rate simultaneously across the P PCs to be coded; i.e., post-compression rate-distortion (PCRD) optimization is applied simultaneously to all code blocks in all PCs to optimally truncate the embedded bit stream for each code block. In the reconstruction process, the Kakadu decoder automatically extracts the transform matrix and mean vector and then applies them appropriately after the bit stream has been decoded (Du 2007).

Due to processing complexity of PCA, a simple algorithm is needed to reduce dimensionality of spectral data. The cubic spline interpolation (CSI) was considered here to reduce the dimensionality of spectral domain of multispectral images (MSIs). The CSI was employed in the spectral domain along with the JPEG that was employed in spatial domain. Down sampling stage before the baseline JPEG was skipped to avoid the loss of sheer information in both the algorithms. The performance results of both algorithms were then compared in terms of compressibility, processing complexity and reliability.

## 3. RESULTS AND DISCUSSION

Four spectral images were selected to test both compression algorithms. Selected four images that include human skin color, natural objects, dark colors and most widely used test color checker chart, are shown below (Figure 1). The images were named as Caucasian male, red rose, tomatoes and Macbeth color checker lab, respectively. Three measures including compression rate (CR), processing time ($T_p$) and CIEDE2000 were used for performance analysis. The CIEDE2000 is CIE color difference formula which is best correlated with the human visual perception (Luo 2001).

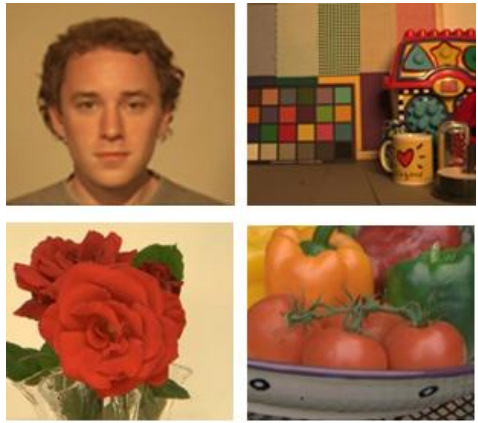

*Figure 1: Selected test images*

The images were calculated using MATLAB. The compression rate was fixed at 8 and performance of both algorithms was tested in terms of quality of visual performance and computational complexity. The average results of image calculations have been shown (see Table 1). The results in (Table 1) show that $\Delta E_{00}$ for PCA based algorithm is much lower as compared to CSI based algorithm. And both algorithms have different performance for different images of same size that is the effect of spatial non-uniformity on the compressibility. The Fig. 2 shows the logarithmic values of processing time of two algorithms at different image sizes. It can be seen from Fig. 2 that increasing the image size processing time of PCA based algorithm increases rapidly, while CSI based algorithm does not add that much computational complexity with the increase of image size.

Being color scientists, we believe that color difference less than 1.0 $\Delta E_{00}$ units is tolerable. Test results showed that for a fixed value of compression rate, CSI based algorithm performed poor in terms of $\Delta E_{00}$, in comparison with PCA, but is still reliable because of small color difference with the original image data. On the other hand it has lower complexity and is computationally much better as compared to PCA based algorithm, especially for spectral images with large number of pixels that may take several minutes to process even on a powerful computer. With the increase of image size, complexity of PCA increases rapidly and hence takes more time to process.

*Table 1. Summary of the results from spectral compression*

| Algo. Name | caucassian | Red Rose | Tomotoes | Macbeth |
|---|---|---|---|---|
| CSI and JPEG | 0.5 | 0.8 | 0.7 | 0.6 |
| PCA and JPEG | 0.2 | 0.3 | 0.2 | 0.2 |

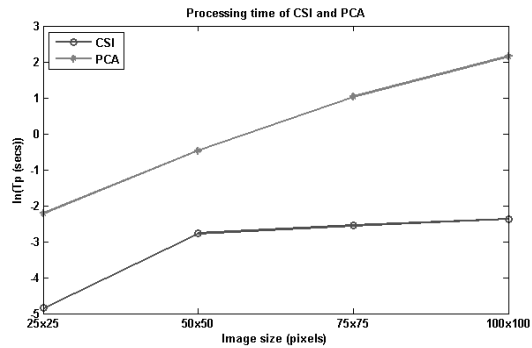

*Figure 2: Processing time plotted against image size.*

## 4. CONCLUSIONS

Two different techniques cubic spline interpolation (CSI) and principle component analysis (PCA) were employed one by one in the spectral domain and were amalgamated with the JPEG, which was employed in spatial domain. The performance results of both algorithms were then compared in terms of compressibility and processing complexity. Test results showed that for a fixed value of compression rate, CSI based algorithm performed poor in terms of $\Delta E_{00}$, in comparison with PCA, but is still reliable because of small color difference with the original image data. On the other hand it has lower complexity and is computationally much better as compared to PCA based algorithm, especially for spectral images with large size. With the increase of image size, complexity of PCA increases rapidly and hence needs more time and memory to process.

*Address: Prof. Ming Ronnier Luo, State Key Laboratory of modern optical instrumentation, Zhejiang University, Hangzhou, China*
*E-mails: msafdar87@zju.edu.cn, m.r.luo@leeds.ac.uk, wangflxy@zju.edu.cn*